\documentclass[aps, prd, preprint, floatfix, nofootinbib, amsmath, amssymb, superscriptaddress]{revtex4}
\usepackage{graphicx}
\usepackage{hyperref}

\newcommand{\Btwiddle}{\widetilde B}
\newcommand{\la}{\lambda}
\newcommand{\lahat}{\hat\lambda}
\newcommand{\muhat}{\hat\mu}

\newcommand{\mean}[1]{\left\langle #1 \right\rangle}
\newcommand{\fc}{f_c = 10.8^{+.10}_{-.05}}

\begin{document}
\title{An improved lattice measurement \\ of the critical coupling in $\phi_2^4$ theory}
\author{David Schaich}
\email{schaich@physics.bu.edu}
\affiliation{Department of Physics and Center for Computational Science, Boston University, Boston, MA 02215}
\affiliation{Department of Physics, Amherst College, Amherst, MA 01002}

\author{Will Loinaz}
\email{waloinaz@amherst.edu}
\affiliation{Department of Physics, Amherst College, Amherst, MA 01002}

\date{30 January 2009}

\begin{abstract} 
  We use Monte Carlo simulations to obtain an improved lattice measurement of the critical coupling constant $\left[\la / \mu^2\right]_{crit}$ for the continuum ($1 + 1$)-dimensional $(\la / 4) \phi^4$ theory.  We find that the critical coupling constant depends logarithmically on the lattice coupling, resulting in a continuum value of $\left[\la / \mu^2\right]_{crit} = 10.8^{+.10}_{-.05}$, in considerable disagreement with the previously reported $\left[\la / \mu^2\right]_{crit} = 10.26^{+.08}_{-.04}$.  Although this logarithmic behavior was not observed in earlier lattice studies, it is consistent with them, and expected analytically.
\end{abstract}

\maketitle

\section{Introduction} 

The two-dimensional $\phi_2^4$ field theory specified by the Euclidean Lagrangian
\begin{equation}
  \label{EQFT}
  \mathcal L_E = \frac{1}{2}(\nabla\phi)^2 + \frac{1}{2}\mu_0^2\phi^2 + \frac{\lambda}{4}\phi^4
\end{equation}
exhibits a phase transition between a symmetric phase with $\mean{\phi} = 0$ and a phase in which the discrete symmetry of the Lagrangian under $\phi \to -\phi$ is broken \cite{chang, glimm}.  Loinaz and Willey \cite{loinaz} have used Monte Carlo simulations to calculate the critical value of the coupling constant that separates the two phases of the theory.

In this work we perform similar calculations, discretizing the Euclidean quantum field theory (EQFT) of Eqn.~\ref{EQFT} in terms of the two dimensionless lattice parameters
\begin{align}
  \lahat & \equiv \la a^2 & \muhat_0^2 & \equiv \mu_0^2a^2,
\end{align}
where $a > 0$ is the lattice spacing.  (In two dimensions, both $\la$ and $\mu_0^2$ have mass dimension $[\la] = \left[\mu_0^2\right] = 2$.)  The lattice action that regularizes Eqn.~\ref{EQFT} is
\begin{equation}
  \label{latticeAction}
  \mathcal A = \sum_n \left[\frac{1}{2}\sum_{\nu=1}^d\left(\phi_{n + e_{\nu}} - \phi_n\right)^2 + \frac{1}{2}\muhat_0^2\phi_n^2 + \frac{\lahat}{4}\phi_n^4\right],
\end{equation}
where $e_{\nu}$ is the unit vector in the $\nu$ direction.  The EQFT is the continuum limit $a \to 0$ of this lattice model.

In two dimensions, the field strength and self-coupling renormalization factors $Z_{\phi}$ and $Z_{\lambda}$ are finite, and do not affect the phase structure of the theory.  However, there is an infinite mass renormalization, which requires that the bare mass parameter be tuned to infinity as the continuum limit is taken, $\mu_0^2 \sim \mu^2 \ln(1 / a)$, where $\mu^2$ is the finite renormalized mass squared.  Since $\la$ is independent of $a$ and $\mu_0^2$ diverges only logarithmically as $a \to 0$, both $\lahat$ and $\muhat_0^2$ vanish in the continuum limit $a \to 0$.  Taking the continuum limit therefore reduces the number of independent dimensionless parameters from two to one, which we take to be the dimensionless coupling constant $f = \la / \mu^2$.

We can parametrize the mass renormalization as
\begin{align}
  \label{additive}  \mu_0^2 & = \mu^2 - \delta\mu^2, \\
               \mathcal L_E & = \frac{1}{2}(\nabla\phi)^2 + \frac{1}{2}\mu^2\phi^2 + \frac{\lambda}{4}\phi^4 - \frac{1}{2}\delta\mu^2\phi^2,
\end{align}
where $\mu^2$ and the finite part of $\delta\mu^2$ depend on the choice of renormalization condition.  We want to choose a renormalization scheme in which the effective coupling constant $f$ distinguishes between the two phases of the theory, which is not the case for several popular renormalization conditions \cite{loinaz}.  We will achieve this by choosing the mass renormalization to be equivalent to normal-ordering the interaction in the interaction picture in the symmetry phase.

\begin{figure}[tp]
    \centering
    \includegraphics{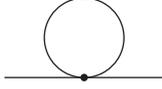}
    \caption{The only divergent Feynman diagram in $\phi^4_2$ theory.}
    \label{diagram}
\end{figure}

There is only one divergent Feynman diagram in $\phi_2^4$ theory, Fig.~\ref{diagram}, which involves the integral
\begin{align}
  \nonumber A_{\mu^2} & = \frac{1}{N^2} \sum_{k_1 = 1}^N\sum_{k_2 = 1}^N\frac{1}{\muhat^2 + 4\sin^2(\pi k_1 / N) + 4\sin^2(\pi k_2 / N)} \\
  \label{infinite}    & \to \int\frac{d^2p}{(2\pi)^2}\frac{1}{p^2 + \mu^2}
\end{align}
in the continuum limit.  From Eqns.~\ref{EQFT} and \ref{additive},
\begin{align}
  G^{-1}(p^2) & = p^2 + \mu_0^2 + \Sigma_0(p^2) = p^2 + \mu^2 + \Sigma(p^2), \\
  \Sigma(p^2) & = 3\la A_{\mu^2} - \delta\mu^2 + \mbox{two-loop}.
\end{align}
Therefore the renormalization condition
\begin{equation}
  \delta\mu^2 = 3\la A_{\mu^2}
\end{equation}
removes all ultraviolet divergence from the perturbation series based on the renormalized parametrization of Eqn.~\ref{additive}.

Applying this renormalization condition,
\begin{align}
  \nonumber \mathcal L_E & = \frac{1}{2}(\nabla\phi)^2 + \frac{1}{2}\mu^2\phi^2 + \frac{\lambda}{4}\phi^4 - \frac{3}{2}\la A_{\mu^2}\phi^2 \\
  \label{ordered}        & = \frac{1}{2}(\nabla\phi)^2 + \frac{1}{2}\mu^2\phi^2 + \frac{\lambda}{4}\mbox{:}\phi^4\mbox{:}_{\mu^2},
\end{align}
dropping a constant piece in the second line.  In terms of $f = \la / \mu^2$, the first line of Eqn.~\ref{ordered} can be written as
\begin{equation}
  \label{couplingL}
  \mathcal L_E = \frac{1}{2}(\nabla\phi)^2 + \frac{1}{2}\mu^2(1 - 3fA_{\mu^2})\phi^2 + \frac{f\mu^2}{4}\phi^4.
\end{equation}

On the lattice ($a > 0$), $A_{\mu^2}$ is finite, so we can argue that for $f$ sufficiently small, the exact effective potential has a single minimum at $\mean{\phi} = 0$.  The coefficient of $\phi^2$ in Eqn.~\ref{couplingL} is negative for large $f$, suggesting a transition to the broken symmetry phase.  However, the effective potential need not be well approximated by its tree-level form at strong coupling.  Chang \cite{chang} has shown that this transition does occur, using a duality transformation from the strong coupling regime of Eqn.~\ref{couplingL} to a weakly-coupled theory normal-ordered with respect to the vacuum of the broken symmetry phase.

We proceed by using Monte Carlo simulations to map the critical line in the $(\muhat_0^2, \lahat)$ plane.  We determine critical values of $\muhat_{0c}^2(\lahat)$ for various $\lahat$, calculating the infinite-lattice-size limit of Monte Carlo data measured on lattices of finite size.  We then impose our renormalization condition
\begin{equation}
  \label{renorm}
  \muhat^2 = \muhat_0^2 + 3\lahat A_{\mu^2}
\end{equation}
using the integral representation of $A_{\mu^2}$ in the infinite-volume limit,
\begin{equation}
  \label{bessel}
  A_{\mu^2} = \int_0^{\infty}dt\exp\left[-\muhat^2t\right]\left(\exp(-2t)I_0(2t)\right)^2.
\end{equation}
Here $I_0$ is a modified Bessel function of the first kind.

For fixed $\lahat \ne 0$, we solve Eqns.~\ref{renorm} and \ref{bessel} numerically to determine $\muhat_c^2$ from $\muhat_{0c}^2$.  We then extrapolate $\lahat \to 0$ to obtain the critical coupling constant
\begin{equation}
  \left[\la / \mu^2\right]_{crit} \equiv f_c = \lim_{\lahat, \muhat^2 \to 0}\left[\lahat / \muhat_c^2\right]
\end{equation}
in the continuum limit.  We will see that this extrapolation has a nonlinear form.

\section{Simulations} 

We performed Monte Carlo simulations based on the lattice action of Eqn.~\ref{latticeAction} on $N\times N$ lattices with $N = 32$, 64, 128, 256, 512, and 1024.  For each $N$, we set $\lahat = 1.0$, 0.7, 0.5, 0.25, 0.1, 0.05, 0.03, 0.02, and 0.01, and for each $(N, \lahat)$ scanned in $\muhat_0^2$ beginning in the symmetric phase and ending in the broken symmetry phase.  To further constrain the data at small $\lahat$, we performed additional simulations at $N = 600$ and 1200 for $\lahat = 0.05$, 0.03, 0.02, and 0.01.

To reduce critical slowing down, our simulations execute a Wolff cluster algorithm \cite{wolff} update on the embedded Ising model after every five random sweeps of the lattice with standard Metropolis updating, as in \cite{brower, loinaz}.  After an initial thermalization of $2^{13}$-$2^{14}$ Metropolis-Wolff cycles, we measured lattice quantities following each of an additional $2^{13}$-$2^{14}$ cycles.

Since these measurements are not independent, we also calculated the autocorrelation time $\tau$ for each $(N, \lahat, \muhat_0^2)$ simulation and incorporated it into our analysis.  Typical autocorrelation times are around ten measurements, with maximum autocorrelation times around 100 measurements for $\lahat \ll 1$ on small lattices.  In every simulation the thermalization time exceeded $100\tau$ and we took at least 100 statistically independent measurements.  As a result, our statistical uncertainties are quite small in comparison to systematic uncertainties.

We use three diagnostics to determine the critical value of $\muhat_{0c}^2$ where the phase transition occurs for fixed $\lahat$.  The first is the familiar peak in the susceptibility $\chi \propto \mean{\phi^2} - \mean{|\phi|}^2$, with uncertainty extracted from the full width of the peak at half its maximum value (FWHM).

\begin{figure}[tp]
    \centering
    \includegraphics{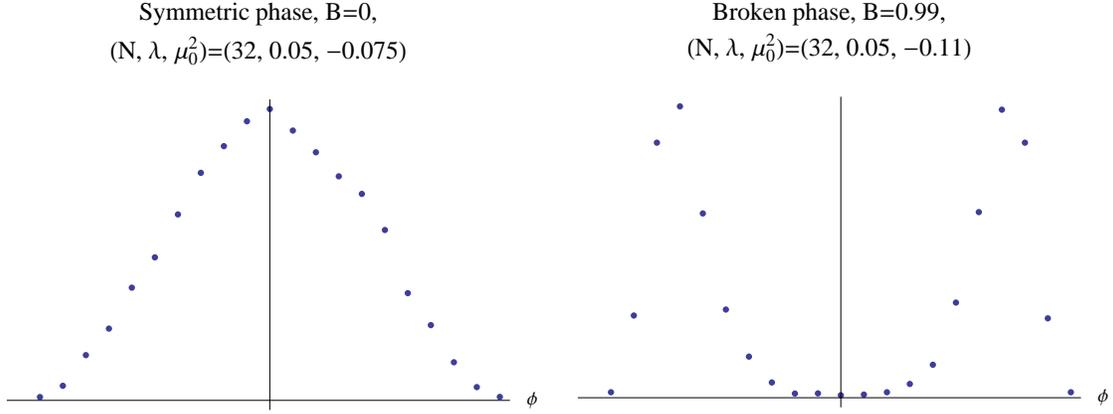}
    \caption{Histograms of $\phi$ for simulations with $N = 32$ and $\lahat = 0.05$, in the symmetric phase (left, $\muhat_0^2 = -0.075$) and broken symmetry phase (right, $\muhat_0^2 = -0.11$).}
    \label{bimod}
\end{figure}

The second diagnostic is the bimodality $B(\muhat_0^2)$, a parameterization of the shape of the histogram of the values of $\phi$ measured during each simulation with fixed $(N, \lahat, \muhat_0^2)$ \cite{loinaz}.  Fig.~\ref{bimod} illustrates these histograms in the two phases of $\phi_2^4$ theory: in the symmetric phase the histogram has a single peak around $\mean{\phi} = 0$, while in the broken symmetry phase it has two peaks, around $\pm\mean{|\phi|} \ne 0$.  Constructing the histogram with an odd number of bins, we define the bimodality as
\begin{equation}
  B = 1 - \frac{n_0}{n_{max}},
\end{equation}
where $n_0$ is the number of measurements in the central bin around zero, and $n_{max}$ is the largest number in any bin.  In the symmetric phase, $B \ll 1$, while in the broken symmetry phase $B \approx 1$ (cf.\ Fig.~\ref{bimod}).

\begin{figure}[tp]
    \centering
    \includegraphics{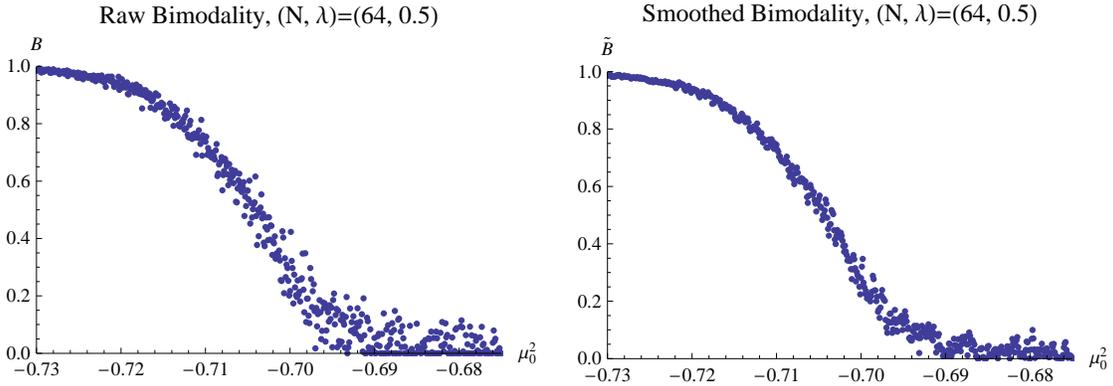}
    \caption{Bimodality plotted against $\muhat_0^2$ for simulations with $N = 64$ and $\lahat = 0.5$, before (left) and after (right) smoothing.}
    \label{smoothing}
\end{figure}

Since $B$ depends on the specific evolution of the system, it can vary considerably for similar values of $\muhat_0^2$, particularly in the symmetric phase.  To smooth out this jitter, we consider the three-point running average $\Btwiddle(\muhat_0^2)$ of $B(\muhat_0^2)$ over $\muhat_0^2$,
\begin{equation}
  \Btwiddle(\muhat_0^2) = \left[B(\muhat_0^2 - \Delta\muhat_0^2) + B(\muhat_0^2) + B(\muhat_0^2 + \Delta\muhat_0^2)\right] / 3.
\end{equation}
Fig.~\ref{smoothing} illustrates the benefits of this smoothing procedure.  We take as the phase transition point the value of $\muhat_0^2$ for which $\Btwiddle(\muhat_0^2)$ is closest to 0.5, with bounds given by the $\muhat_0^2$ most distant from this critical $\muhat_{0c}^2$ for which $0.1 < \Btwiddle < 0.95$ (cf.\ Fig.~\ref{smoothing}).  These conventions produce results consistent with those from the susceptibility, with comparable (though generally smaller) uncertainties, as shown in Table~\ref{indicators}.

To verify that the bimodality is a robust indicator of the phase transition, we checked its behavior in the well-understood two-dimensional Ising model.  Using the conventions stated above, we found that the critical $\muhat_{0c}^2$ indicated by the bimodality agrees well with that indicated by the susceptibility in this case as well, with comparable uncertainties.  Both observables agree with the exact analytic result.

Finally, we extract a third estimate of the critical $\muhat_{0c}^2$ using the Binder cumulant \cite{binder}
\begin{equation}
  \label{cumulant}
  U = 1 - \frac{\mean{\phi^4}}{3\mean{\phi^2}^2}.
\end{equation}
For $\lahat$ fixed, $U$ has a fixed point at the critical $\muhat_{0c}^2$ for any value of the lattice size $N$.  We take as the critical $\muhat_{0c}^2$ the value of $\muhat_0^2$ at which $U$ for the three largest $N$ are closest together, with bounds given by the $\muhat_0^2$ at which all three separate.

This analysis of the cumulant in Eqn.~\ref{cumulant} produces a single critical $\muhat_{0c}^2$ for each $\lahat$, while we have susceptibility and bimodality data for each ($N$, $\lahat$).  Performing a linear regression to find the $N \to \infty$ limit of the susceptibility and bimodality data with $\lahat$ fixed gives us a total of three independent indicators of the critical $\muhat_{0c}^2$ for each $\lahat$.

We find all three values for each $\lahat$ consistent with each other, with comparable uncertainties (Table~\ref{indicators}).  Combining them produces the second column in Table~\ref{lattice_results}.  The third column  in Table~\ref{lattice_results} holds the corresponding critical renormalized $\muhat_c^2$ determined from Eqns.~\ref{renorm} and \ref{bessel}, while the fourth presents the values of the critical coupling $\lahat / \muhat_c^2$ which are to be extrapolated to the $a \to 0$ continuum limit.

\begin{table}[tp]
  \caption{Critical $\muhat_{0c}^2$ from each phase transition indicator.}
  \centering
  \begin{ruledtabular}
  \begin{tabular}{cccc}
    $\lahat$  & Susceptibility  & Bimodality    & Cumulant      \\
    \hline
    1.00      & -1.27233(16)    & -1.27258(10)  & -1.27260(45)  \\
    0.70      & -0.95151(25)    & -0.95152(7)   & -0.95180(40)  \\
    0.50      & -0.72080(11)    & -0.72131(9)   & -0.72130(30)  \\
    0.25      & -0.40346(18)    & -0.40373(6)   & -0.40390(20)  \\
    0.10      & -0.18424(11)    & -0.18432(9)   & -0.18430(20)  \\
    0.05      & -0.10060(5)     & -0.10071(4)   & -0.10100(35)  \\
    0.03      & -0.06410(4)     & -0.06414(5)   & -0.06420(15)  \\
    0.02      & -0.04464(3)     & -0.04468(5)   & -0.04500(30)  \\
    0.01      & -0.02397(6)     & -0.02399(5)   & -0.02410(10)  \\
  \end{tabular}
  \end{ruledtabular}
  \label{indicators}
\end{table}

\begin{table}[tp]
  \caption{Critical $\muhat_{0c}^2$, $\muhat_c^2$ and $\lahat / \muhat_c^2$ for different $\lahat$.}
  \centering
  \begin{ruledtabular}
  \begin{tabular}{cccc}
    $\lahat$  & $\muhat_{0c}^2$ & $\muhat_c^2$  & $\lahat / \muhat_c^2$ \\
    \hline
    1.00      & -1.27251(16)    & 0.097320(46)  & 10.275(5)             \\
    0.70      & -0.95153(16)    & 0.068462(45)  & 10.225(7)             \\
    0.50      & -0.72112(11)    & 0.048884(32)  & 10.228(7)             \\
    0.25      & -0.40372(9)     & 0.024176(26)  & 10.341(11)            \\
    0.10      & -0.18429(8)     & 0.009476(23)  & 10.553(26)            \\
    0.05      & -0.10067(12)    & 0.004679(33)  & 10.686(76)            \\
    0.03      & -0.06412(5)     & 0.002794(15)  & 10.737(59)            \\
    0.02      & -0.04466(10)    & 0.001870(28)  & 10.695(163)           \\
    0.01      & -0.02400(4)     & 0.000931(12)  & 10.739(138)           \\
  \end{tabular}
  \end{ruledtabular}
  \label{lattice_results}
\end{table}

\section{Analysis} 

\begin{figure}[tp]
    \centering
    \includegraphics{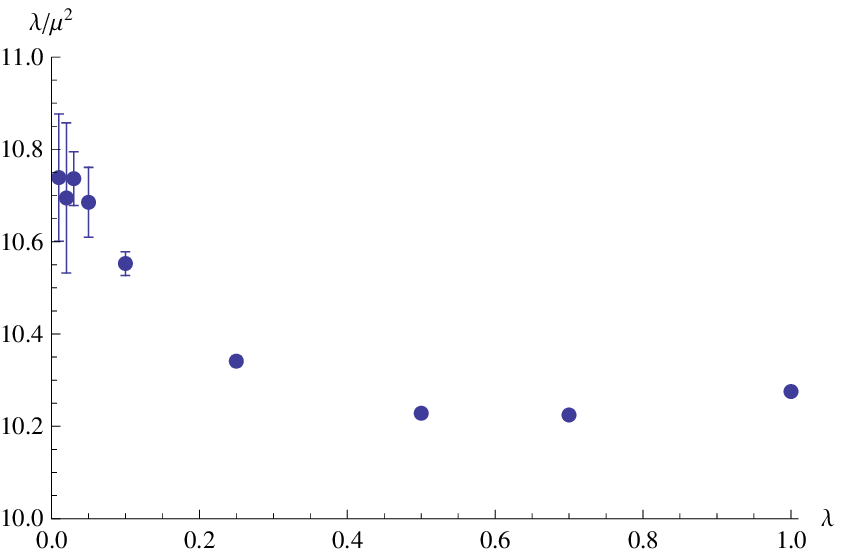}
    \caption{Critical coupling constant $\lahat / \muhat_c^2$ plotted against $\lahat$.}
    \label{data_points}
\end{figure}

Fig.~\ref{data_points} plots the values of $\lahat / \muhat_c^2$ in the fourth column of Table~\ref{lattice_results} and clearly rules out a linear $\lahat \to 0$ extrapolation like that performed in \cite{loinaz}.

Analytic investigations into the structure of scalar field theories, and super-renormalizable theories more generally, long ago established that correlation functions in these theories typically depend on logarithms of the coupling \cite{parisi, symanzik-NC, symanzik, jackiw-templeton, jackiw-thesis}.  Jackiw and Templeton \cite{jackiw-templeton} explicitly demonstrated the presence of such logarithmic terms in a simple $\phi_4^3$ model, using a truncated Bethe-Salpeter equation.  This non-analytic dependence on the coupling appears generically in more complicated super-renormalizable theories as well, including the $\phi_2^4$ theory we consider here.

We can numerically investigate the effect of such logarithmic dependence by fitting the data in Table~\ref{lattice_results} and Fig.~\ref{data_points} to a function of the form
\begin{equation}
  \label{directFit}
  \lahat / \muhat_c^2 = c_0 + c_1\lahat + c_2\lahat\ln\lahat.
\end{equation}
The constant $c_0$ is exactly the continuum critical coupling constant $f_c$ we wish to determine.  Performing this fit, we find $c_0 = f_c = 10.78(3)$, with $\chi^2 = 1.21$ per degree of freedom ($dof$).  Performing fits with additional terms ($c_3\lahat^2$ or $c_3\lahat^2\ln\lahat$) results in even larger $f_c \approx 10.9$ with very small $\chi^2/dof \approx 0.15$ (Table~\ref{fits-1}).  Fits that do not include a term logarithmic in $\lahat$ are poor, with $\chi^2/dof \gg 1$.

\begin{table}[tp]
  \caption{$\lahat, \muhat^2 \to 0$ extrapolations of $\lahat / \muhat_c^2$ vs. $\lahat$.}
  \centering
  \begin{ruledtabular}
  \begin{tabular}{ccc}
    Form of $\lahat / \muhat_c^2$ fit                             & $f_c$     & $\chi^2/dof$  \\
    \hline
    $f_c + c_1\lahat$                                             & 10.31(6)  & 48            \\
    $f_c + c_1\lahat + c_2\lahat^2$                               & 10.60(5)  & 5.8           \\
    $f_c + c_1\lahat + c_2\lahat\ln\lahat$                        & 10.78(3)  & 1.2           \\
    $f_c + c_1\lahat + c_2\lahat\ln\lahat + c_3\lahat^2$          & 10.89(2)  & 0.16          \\
    $f_c + c_1\lahat + c_2\lahat\ln\lahat + c_3\lahat^2\ln\lahat$ & 10.87(2)  & 0.13          \\
  \end{tabular}
  \end{ruledtabular}
  \label{fits-1}
\end{table}

We can check the consistency of these results by fitting $\muhat_c^2$ as a function of $\lahat$ and extracting $f_c$ from the coefficient of the term linear in $\lahat$,
\begin{equation*}
  \muhat_c^2 = d_0 + \lahat / f_c + \mathcal O(\lahat^2).
\end{equation*}
Fitting
\begin{equation}
  \label{goodFit}
  \muhat_c^2 = d_0 + \lahat / f_c + d_1\lahat^2 + d_2\lahat^2\ln\lahat,
\end{equation}
we find $f_c = 10.77(4)$ with $\chi^2/dof = 1.1$.  As above, including additional terms in the fit raises $f_c$ while dramatically lowering the $\chi^2/dof$, while fits without any logarithmic term have $\chi^2/dof \gg 1$ (Table~\ref{fits-2}).

Since $\muhat_c^2 \to 0$ as $\lahat \to 0$, we should find the constant term $d_0 \approx 0$ in these fits, and we can also perform fits with $d_0$ explicitly set to zero as an additional check.  Of the fits listed in Table~\ref{fits-2}, only that of Eqn.~\ref{goodFit} has $d_0$ vanish within uncertainty, although $d_0$ is within 2$\sigma$ of zero for the fit form
\begin{equation*}
  \muhat_c^2 = d_0 + \lahat / f_c + d_1\lahat^2 + d_2\lahat^3
\end{equation*}
as well.  The value of $f_c$ extracted from the fit
\begin{equation}
  \muhat_c^2 = \lahat / f_c + d_1\lahat^2 + d_2\lahat^2\ln\lahat,
\end{equation}
is $f_c = 10.79(3)$ with $\chi^2/dof = 1.0$, in agreement with the values from Eqns.~\ref{directFit} and \ref{goodFit}.

\begin{table}[tp]
  \caption{$\lahat \to 0$ extrapolations of $\muhat_c^2$ vs. $\lahat$.}
  \centering
  \begin{ruledtabular}
  \begin{tabular}{ccc}
    Form of $\muhat_c^2$ fit                                                          & $f_c$     & $\chi^2/dof$  \\
    \hline
    $d_0 + \lahat / f_c$                                                              & 10.24(2)  & 28            \\
    $d_0 + \lahat / f_c + d_1\lahat^2$                                                & 10.24(6)  & 33            \\
    $d_0 + \lahat / f_c + d_1\lahat^2 + d_2\lahat^3$                                  & 10.55(5)  & 4.1           \\
    $d_0 + \lahat / f_c + d_1\lahat^2 + d_2\lahat^2\ln\lahat$                         & 10.77(4)  & 1.1           \\
    $d_0 + \lahat / f_c + d_1\lahat^2 + d_2\lahat^2\ln\lahat + d_3\lahat^3$           & 10.98(2)  & 0.04          \\
    $d_0 + \lahat / f_c + d_1\lahat^2 + d_2\lahat^2\ln\lahat + d_3\lahat^3\ln\lahat$  & 10.93(2)  & 0.05          \\
    $\lahat / f_c$                                                                    & 10.27(2)  & 47            \\
    $\lahat / f_c + d_1\lahat^2$                                                      & 10.31(7)  & 49            \\
    $\lahat / f_c + d_1\lahat^2 + d_2\lahat^3$                                        & 10.61(5)  & 5.5           \\
    $\lahat / f_c + d_1\lahat^2 + d_2\lahat^2\ln\lahat$                               & 10.79(3)  & 1.0           \\
    $\lahat / f_c + d_1\lahat^2 + d_2\lahat^2\ln\lahat + d_3\lahat^3$                 & 10.90(2)  & 0.18          \\
    $\lahat / f_c + d_1\lahat^2 + d_2\lahat^2\ln\lahat + d_3\lahat^3\ln\lahat$        & 10.89(2)  & 0.14          \\
  \end{tabular}
  \end{ruledtabular}
  \label{fits-2}
\end{table}

Clearly, systematic errors, particularly the choice of continuum extrapolation form, dominate over statistical errors.  Tables~\ref{fits-1} and \ref{fits-2} summarize $f_c$ for various linear and nonlinear extrapolations, along with the goodness of the fits, $\chi^2/dof$.  Neglecting fits with $\chi^2/dof \gg 1$, we adopt a final result of
\begin{equation}
  \fc
\end{equation}
to be consistent with the numbers in Tables~\ref{fits-1} and \ref{fits-2}.

\section{Discussion} 

\begin{figure}[tp]
    \centering
    \includegraphics{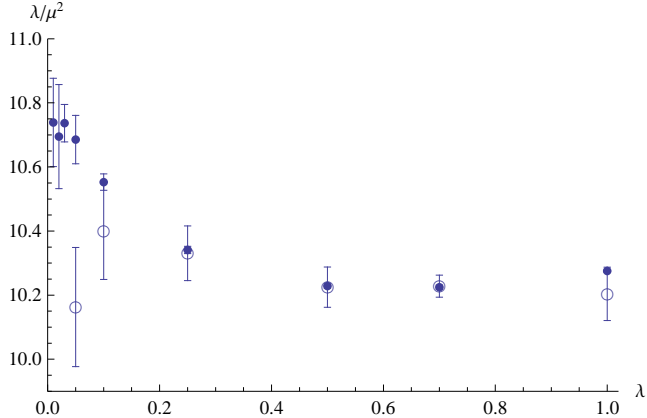}
    \caption{Our data for $\lahat / \muhat_c^2$ compared with the results of \cite{loinaz} (empty circles).}
    \label{compare}
\end{figure}

Since our approach closely parallels that of Loinaz and Willey \cite{loinaz}, it is distressing that our final result disagrees so strongly with the $f_c = 10.26^{+.08}_{-.04}$ reported there.  However, our individual data points are largely consistent with theirs, as shown in Fig.~\ref{compare}.  The disagreement between our final results comes almost entirely from the nonlinear continuum extrapolations discussed above.

Re-analyzing the data in \cite{loinaz}, we find (Tables~\ref{1997fits-1} and \ref{1997fits-2}) that they are consistent with all the nonlinear fits considered in our analysis above.  Both linear and nonlinear fits all have $\chi^2/dof \sim 0.5$.  While nonlinear extrapolations were not required by the data in \cite{loinaz}, by considering only the linear case Loinaz and Willey overlooked significant systematic effects due to fit form.  In Tables~\ref{1997fits-1} and \ref{1997fits-2}, we see $10.2 \lesssim f_c \lesssim 11.9$ in fits with $\chi^2/dof < 1$, consistent with our result $\fc$.

\begin{table}[tp]
  \caption{$\lahat, \muhat^2 \to 0$ extrapolations of $\lahat / \muhat_c^2$ vs. $\lahat$, for \cite{loinaz}.}
  \centering
  \begin{ruledtabular}
  \begin{tabular}{ccc}
    Form of $\lahat / \muhat_c^2$ fit                             & $f_c$     & $\chi^2/dof$  \\
    \hline
    $f_c + c_1\lahat$                                             & 10.32(5)  & 0.39          \\
    $f_c + c_1\lahat + c_2\lahat^2$                               & 10.34(9)  & 0.49          \\
    $f_c + c_1\lahat + c_2\lahat\ln\lahat$                        & 10.34(12) & 0.51          \\
    $f_c + c_1\lahat + c_2\lahat\ln\lahat + c_3\lahat^2$          & 10.18(25) & 0.59          \\
    $f_c + c_1\lahat + c_2\lahat\ln\lahat + c_3\lahat^2\ln\lahat$ & 10.23(22) & 0.63          \\
  \end{tabular}
  \end{ruledtabular}
  \label{1997fits-1}
\end{table}

\begin{table}[tp]
  \caption{$\lahat \to 0$ extrapolations of $\muhat_c^2$ vs. $\lahat$, for \cite{loinaz}.}
  \centering
  \begin{ruledtabular}
  \begin{tabular}{ccc}
    Form of $\muhat_c^2$ fit                                                          & $f_c$     & $\chi^2/dof$  \\
    \hline
    $d_0 + \lahat / f_c$                                                              & 10.23(3)  & 0.65          \\
    $d_0 + \lahat / f_c + d_1\lahat^2$                                                & 10.37(9)  & 0.44          \\
    $d_0 + \lahat / f_c + d_1\lahat^2 + d_2\lahat^3$                                  & 10.57(16) & 0.32          \\
    $d_0 + \lahat / f_c + d_1\lahat^2 + d_2\lahat^2\ln\lahat$                         & 10.80(24) & 0.23          \\
    $d_0 + \lahat / f_c + d_1\lahat^2 + d_2\lahat^2\ln\lahat + d_3\lahat^3$           & 11.86(23) & 0.02          \\
    $d_0 + \lahat / f_c + d_1\lahat^2 + d_2\lahat^2\ln\lahat + d_3\lahat^3\ln\lahat$  & 11.64(12) & 0.01          \\
    $\lahat / f_c$                                                                    & 10.24(2)  & 0.57          \\
    $\lahat / f_c + d_1\lahat^2$                                                      & 10.32(5)  & 0.38          \\
    $\lahat / f_c + d_1\lahat^2 + d_2\lahat^3$                                        & 10.35(9)  & 0.49          \\
    $\lahat / f_c + d_1\lahat^2 + d_2\lahat^2\ln\lahat$                               & 10.35(12) & 0.50          \\
    $\lahat / f_c + d_1\lahat^2 + d_2\lahat^2\ln\lahat + d_3\lahat^3$                 & 10.19(25) & 0.59          \\
    $\lahat / f_c + d_1\lahat^2 + d_2\lahat^2\ln\lahat + d_3\lahat^3\ln\lahat$        & 10.24(22) & 0.63          \\
  \end{tabular}
  \end{ruledtabular}
  \label{1997fits-2}
\end{table}

\begin{table}[tp]
  \caption{Various results for the critical coupling $f_c$.}
  \centering
  \begin{ruledtabular}
  \begin{tabular}{ccc}
    Method                                & Result                & Reference             \\
    \hline
    Monte Carlo                           & $10.8^{+.10}_{-.05}$  & This work             \\
    Gaussian effective potential          & 10.272                & \cite{hauser}         \\
    Gaussian effective potential          & 10.211                & \cite{chang}          \\
    GEP and oscillator rep.               & 10.21                 & \cite{vinnikov}       \\
    Spherical field theory                & 10.05                 & \cite{lee}            \\
    Diffusion Monte Carlo                 & $10 \pm 0.8 \pm 0.4$  & \cite{marrero}        \\
    Density matrix RG                     & 9.982(2)              & \cite{sugihara}       \\
    Continuum light-front                 & 9.91                  & \cite{bender}         \\
    Connected Green function              & 9.784                 & \cite{hauser}         \\
    Coupled cluster expansion             & $3.80 < f_c < 8.60$   & \cite{funke}          \\
    Discretized light-front               & 7.325, 7.71           & \cite{sugihara-older} \\
    Discretized light-front               & 7.316, 5.500          & \cite{har-1, har-2}   \\
    Random phase approximation            & 7.2                   & \cite{hansen}         \\
    Non-Gaussian variational              & 6.88                  & \cite{polley}         \\
  \end{tabular}
  \end{ruledtabular}
  \label{methods}
\end{table}

Several other authors have also calculated the critical coupling $f_c$ in $\phi_2^4$ theory using a variety of methods, numerical schemes, and analytic approximations.  These approaches, summarized in Table~\ref{methods}, produce a large spread of results, of which ours is the largest.

The density matrix renormalization group result $f_c = 9.982(2)$ \cite{sugihara} is notable for its extremely small claimed uncertainty.  However, this result relies on linear $\lahat \to 0$ extrapolations like those in \cite{loinaz}, performed with just two data points at $\lahat = 0.1$, 0.25, for fixed lattice size $N = 500$ or 1000.  A linear $1 / N \to 0$ extrapolation is then performed using the two resulting values.  Thus, we expect this result to suffer from the difficulties discussed above.

The diffusion Monte Carlo result $f_c = 10 \pm 0.8 \pm 0.4$ \cite{marrero} agrees with our result within its relatively large statistical and systematic uncertainties.  The Gaussian effective potential results $f_c = 10.272$ \cite{hauser}, $f_c = 10.211$ \cite{chang}, and $f_c = 10.21$ \cite{vinnikov} (the last of which coincides with the oscillator representation result) are the next closest.  Both the Gaussian effective potential and oscillator representation methods incorrectly predict a first-order phase transition, in violation of the Simon-Griffiths theorem \cite{simon}, which requires the $\phi_2^4$ theory phase transition to be second order.

\section{Conclusions} 

We have used Monte Carlo simulations to obtain an accurate lattice measurement of the continuum critical coupling constant $\fc$ in $\phi_2^4$ theory, improving the previously reported Monte Carlo result \cite{loinaz}.

While our data are consistent with those reported in \cite{loinaz}, our improved precision forces a nonlinear extrapolation to the continuum limit, producing a significantly different continuum result.  The older data are compatible with these nonlinear extrapolations, although such nonlinearity was neither required nor investigated previously.  Applying nonlinear extrapolations to the older data, we obtain continuum results consistent with our own.

Significantly, nonlinearity -- in particular terms logarithmic in the lattice coupling $\lahat$ -- is expected analytically.  This convergence of analytic theory and numerical data provides additional evidence that our improved result is accurate and reliable, and can be used to evaluate analytic approximations.

\begin{acknowledgments} 
This work was supported by the National Science Foundation under Grants No.\ CNS-0521169 and No.\ DGE-0221680.
\end{acknowledgments}

\raggedright
\bibliographystyle{plain}

\begin{thebibliography}{99}
\bibitem{chang}
  S.~J.~Chang,
  \href{http://dx.doi.org/10.1103/PhysRevD.13.2778}{Phys.\ Rev.\ {\bf D13}:2778-2788} (1976)
  [Erratum-\href{http://dx.doi.org/10.1103/PhysRevD.16.1979}{ibid.\ {\bf D16}:1979} (1977)].

\bibitem{glimm}
  J.~Glimm and A.~Jaffe.
  \emph{Quantum Physics: A Functional Integral Point of View}.
  Springer-Verlag, New York, 1981.

\bibitem{loinaz}
  Will Loinaz and R.~S.~Willey,
  \href{http://dx.doi.org/10.1103/PhysRevD.58.076003}{Phys.\ Rev.\ {\bf D58}:076003} (1998).

\bibitem{wolff}
  U.~Wolff,
  \href{http://dx.doi.org/10.1103/PhysRevLett.62.361}{Phys.\ Rev.\ Lett.\ {\bf 62}:361-364} (1989).

\bibitem{brower}
  R.~C.~Brower and P.~Tamayo,
  \href{http://dx.doi.org/10.1103/PhysRevLett.62.1087}{Phys.\ Rev.\ Lett.\ {\bf 62}:1087-1090} (1989).

\bibitem{binder}
  K.~Binder,
  \href{http://dx.doi.org/10.1007/BF01293604}{Z.\ Phys.\ {\bf B43}:119} (1981).

\bibitem{parisi}
  G.~Parisi,
  \href{http://dx.doi.org/10.1016/0550-3213(85)90211-1}{Nucl.\ Phys.\ {\bf B254}:58-70} (1985).

\bibitem{symanzik-NC}
  K.~Symanzik,
  \href{http://dx.doi.org/10.1007/BF02725853}{Lett.\ Nuovo Cim.\ {\bf 8S2}:771} (1973).

\bibitem{symanzik}
  K.~Symanzik,
  DESY preprint 73/58 (1973).

\bibitem{jackiw-templeton}
  R.~Jackiw and S.~Templeton,
  \href{http://dx.doi.org/10.1103/PhysRevD.23.2291}{Phys.\ Rev.\ {\bf D23}:2291-2304} (1981).

\bibitem{jackiw-thesis}
  R.~Jackiw,
  Ph.D.\ thesis,
  Cornell University, 1966 (unpublished).

\bibitem{hauser}
  J.~M.~Hauser, W.~Cassing, A.~Peter and M.~H.~Thoma,
  \href{http://dx.doi.org/10.1007/BF01292336}{Z.\ Phys.\ {\bf A353}:301-310} (1996).

\bibitem{vinnikov}
  C.~R.~Ji, J.~I.~Kim, D.~P.~Min and A.~V.~Vinnikov,
  \href{http://arxiv.org/abs/hep-ph/0204114}{arXiv:hep-ph/0204114}.

\bibitem{lee}
  D.~Lee,
  \href{http://dx.doi.org/10.1016/S0370-2693(98)01010-7}{Phys.\ Lett.\ {\bf B439}:85-94} (1998).

\bibitem{marrero}
  P.~J.~Marrero, E.~A.~Roura and D.~Lee,
  \href{http://dx.doi.org/10.1016/S0370-2693(99)01341-6}{Phys.\ Lett.\ {\bf B471}:45-52} (1999).

\bibitem{sugihara}
  T.~Sugihara,
  \href{http://dx.doi.org/10.1088/1126-6708/2004/05/007}{JHEP {\bf 0405}:007} (2004).

\bibitem{bender}
  C.~M.~Bender, S.~Pinsky and B.~Van de Sande,
  \href{http://dx.doi.org/10.1103/PhysRevD.48.816}{Phys.\ Rev.\ {\bf D48}:816-821} (1993).

\bibitem{funke}
  M.~Funke, U.~Kaulfuss and H.~Kummel,
  \href{http://dx.doi.org/10.1103/PhysRevD.35.621}{Phys.\ Rev.\ {\bf D35}:621-630} (1987).

\bibitem{sugihara-older}
  T.~Sugihara,
  \href{http://dx.doi.org/10.1103/PhysRevD.57.7373}{Phys.\ Rev.\ {\bf D57}:7373-7382} (1998).

\bibitem{har-1}
  A.~Harindranath and J.~P.~Vary,
  \href{http://dx.doi.org/10.1103/PhysRevD.36.1141}{Phys.\ Rev.\ {\bf D36}:1141-1147} (1987).

\bibitem{har-2}
  A.~Harindranath and J.~P.~Vary,
  \href{http://dx.doi.org/10.1103/PhysRevD.37.1076}{Phys.\ Rev.\ {\bf D37}:1076-1078} (1988).

\bibitem{hansen}
  H.~Hansen, G.~Chanfray, D.~Davesne and P.~Schuck,
  \href{http://dx.doi.org/10.1140/epja/i2002-10023-y}{Eur.\ Phys.\ J.\ {\bf A14}:397-411} (2002).

\bibitem{polley}
  L.~Polley and U.~Ritschel,
  \href{http://dx.doi.org/10.1016/0370-2693(89)90189-5}{Phys.\ Lett.\ {\bf B221}:44-48} (1989).

\bibitem{simon}
  B.~Simon and R.~B.~Griffiths,
  \href{http://dx.doi.org/10.1007/BF01645626}{Commun.\ Math.\ Phys.\ {\bf 33}:145-164} (1973).
\end{thebibliography}


\end{document}